\documentclass[conference]{IEEEtran}
\IEEEoverridecommandlockouts
\usepackage{nicematrix}
\usepackage{enumitem}
\usepackage{color}
\usepackage{cite}
\usepackage{textcomp}
\usepackage{xcolor}
\usepackage{balance}
\usepackage{graphicx}
\usepackage[english]{babel}
\usepackage{amsthm,amssymb}
\usepackage{amsmath}
\usepackage{mathtools}
\usepackage{tikz}
\usepackage{pgfplots}
\usepackage{pgf}
\usepackage{amsfonts}
\usepackage{bm}
\usepackage[bottom]{footmisc}
\usepackage[ruled,linesnumbered]{algorithm2e}
\usepackage[T1]{fontenc}
\usepackage{bbm}
\usepackage{enumerate}

\usepackage{setspace}
\usepackage{array}
\usepackage{multirow}
\usepackage{diagbox}

\newtheorem{lemma}{Lemma}

\allowdisplaybreaks
\usepackage{bbm}
\usepackage{subcaption}
\usepackage{cite}
\usepackage{amsmath,amssymb,amsfonts}
\usepackage{algorithmic}
\usepackage{graphicx}
\usepackage{textcomp}
\usepackage{xcolor}
\usepackage{todonotes}

\def\BibTeX{{\rm B\kern-.05em{\sc i\kern-.025em b}\kern-.08em
    T\kern-.1667em\lower.7ex\hbox{E}\kern-.125emX}}
\begin{document}

\title{User-UAV Association for Dynamic User in mmWave Communication for eMBB and URLLC}

\author{\IEEEauthorblockN{Siddhanta Parial, Sasthi C. Ghosh}
\IEEEauthorblockA{\textit{Advanced Computing \& Microelectronics Unit} \\
\textit{Indian Statistical Institute, Kolkata 700108, India}\\
parial.sid\_r@isical.ac.in, sasthi@isical.ac.in}
\and
\IEEEauthorblockN{Anil K. Ghosh}
\IEEEauthorblockA{\textit{Theoretical Statistics \& Mathematics Unit} \\
\textit{Indian Statistical Institute, Kolkata 700108, India}\\
akghosh@isical.ac.in}
}

\maketitle

\begin{abstract}
In unmanned aerial vehicle (UAV) assisted millimeter wave (mmWave) communication, appropriate user-UAV association is crucial for improving system performance. In mmWave communication, user throughput largely depends on the line of sight (LoS) connectivity with the UAV, which in turn depends on the mobility pattern of the users. Moreover, different traffic types like enhanced mobile broadband (eMBB) and ultra reliable low latency communication (URLLC) may require different types of LoS connectivity. Existing user-UAV association policies do not consider the user mobility during a time interval and different LoS requirements of different traffic types. In this paper, we consider both of them and develop a user association policy in the presence of building blockages. First, considering a simplified scenario, we have analytically established the LoS area, which is the region where users will experience seamless LoS connectivity for eMBB traffic, and the LoS radius, which is the radius of the largest circle within which the user gets uninterrupted LoS services for URLLC traffic. Then, for a more complex scenario, we present a geometric shadow polygon-based method to compute LoS area and LoS radius. Finally, we associate eMBB and URLLC users, with the UAVs from which they get the maximum average throughput based on LoS area and maximum LoS radius respectively. We show that our approach outperforms the existing discretization based and maximum throughput based approaches.
\end{abstract}


\section{Introduction}
\noindent Global $5$G subscriptions are anticipated to reach approximately $6.3$ billion in $2030$ \cite{ericsson}, necessitating a variety of applications with different requirements for energy efficiency, reliability, latency, and data throughput. To fulfill these requirements, international telecommunication union (ITU) has classified $5$G traffic into following three categories: enhanced mobile broad band (eMBB), ultra reliable low latency communication (URLLC) and massive machine type communication (mMTC) \cite{navarro2020survey}.  For eMBB users, maximizing the average throughput is of primary importance as it can tolerate slight delays. Whereas, for URLLC traffic uninterrupted LoS connectivity is required to achieve high reliability and low latency. In mMTC services, connection density of machine-type terminals is of main concern. In this manuscript, we consider eMBB and URLLC that are adequate for human-type cellular communications. While millimeter wave (mmWave) can reliably offer a very high data rate over short distances, line of sight (LoS) between the transmitter and the receiver is required due to its high propagation loss over distances and severe penetration loss from obstacles \cite{bai2014analysis}. Because of the flexibility and ease of deployment, unmanned aerial vehicles (UAVs) can significantly aid in avoiding obstructions and thereby improve LoS connection \cite{li2018uav}. However, to achieve such improvement, a user must be associated with an appropriate UAV.

The issue of user-UAV association is therefore crucial in order to provide the user with the optimal UAV based on various requirements. A joint user association, beamforming design and trajectory planning problem is investigated in \cite{zhang2024joint} to maximize the sum weighted bit rate of all users. The authors in \cite{singh2022user} formulated the user association problem using a restless multi-armed bandit. Users have been associated with UAVs maximizing sum-throughput of users in \cite{mahmood2023joint}. The authors in \cite{van2022urllc} solved user association for URLLC users based on latency minimization with digital twin approach. In \cite{xi2020network}, eMBB users are associated based on payload communication model and URLLC users are served based on control information communication model. A hierarchical UAV-assisted framework has been used to solve the mixed traffic user association problem in \cite{tian2023demand}. In \cite{wang2020trajectory}, the user association problem is solved that maximizes sum log-rate at each time interval. The user association problem is formulated as a cooperative stochastic game in \cite{cheng2024learning} and as a Markov decision process in \cite{guan2021user}. However, most existing studies solve the user association problem based on the position of the users at the beginning of each time interval and assume that the users do not change their position during that interval.
 Consequently, they do not take into account the fact that a user having LoS with the UAV at the beginning and end of that time interval does not necessarily indicate LoS connectivity throughout that entire period. This may lead to a decrease in user throughput depending on the LoS connectivity requirements of different traffic types. 

In this manuscript, we develop a user-UAV association policy considering both the user mobility during a time interval $\Delta t$ and the LoS requirements of different traffic types. To account for user mobility, we assume that during $\Delta t$ duration, the user $g$ moves along a straight line with a constant speed $v_g$ and resides within a disk of radius $v_g\Delta t$ centering the user location at the beginning of $\Delta t$. To account for the LoS requirement of eMBB traffic, the {\it LoS area} is defined as the area inside the mobility region within which the user gets LoS services. Whereas, for URLLC traffic, the {\it LoS radius} is defined as the radius of the largest circle within which the user gets uninterrupted LoS services in its all possible movement directions (Fig \ref{fig:environment}). The computation of LoS area and LoS radius and the subsequent derivation of the association policy can be summarized as follows: 
 \begin{itemize}
 \item Assuming links are independent,
we develop an analytical method for computing eMBB LoS area and URLLC LoS radius based on the building density and LoS probability.
 
\item For more realistic scenario, we develop a shadow polygon aided geometric approach, which computes eMBB LoS area and URLLC LoS radius based on building shadow on the ground with respect to the UAV. 

 \item We derive an association policy where a user is associated to the UAV that offers the maximum average throughput based on LoS area for eMBB and the maximum LoS radius for URLLC.
\end{itemize}

We perform extensive simulations and compare our analytical as well as shadow polygon approaches with  the discretization based method \cite{sabzehali20213d, li2022geometric}. Moreover, we show that our association policy outperforms the existing maximum throughput based policies \cite{wang2020trajectory,mahmood2023joint,zhang2024joint}. Rest of the paper is organized as follows. In section \ref{system model}, the system model is discussed. We compute the eMBB LoS area and the URLLC LoS radius using analytical approach in section \ref{analytical}. The shadow polygon approach and the association policy is described in section \ref{spa}. The simulation results are discussed in section \ref{simulation} and section \ref{conclusion} draws conclusions.

\section{System Model}\label{system model}
\noindent We explore a wireless communication scenario where UAVs are distributed according to homogeneous Poisson point process (PPP) with density $\lambda_U$ in the region. Here we show the computation for one user and the same can be done for all other users. Let the position of the ground user be $g(x_g,y_g,h_g)$ and the position of the $k^{th}$ UAV be $k(x_{k},y_{k},h_{k})$. 

\subsection{User Mobility Model}\label{mob_model}
\noindent The total communication time $t$ is discretized into $n$ number of slots with slot duration $\Delta t$. For a small time duration $\Delta t$, we suppose that the users move along a straight line with a fixed speed $v_g$ taken from uniform $\mathcal{U}\;[0,v_{\rm max}]$, where $v_{\rm max}$ is the maximum speed of any user. So for $\Delta t$ time, the user mobility region will be a disk of radius $R_g=v_g\Delta t$ centering the user location. Since $\Delta t$ is sufficiently small, it is safe to assume that the UAV will remain fixed during the $\Delta t$ interval.
\vspace{-2mm}
\begin{figure}
    \centering
\begin{subfigure}[b]{0.25\textwidth}  
    \includegraphics[width=\linewidth]{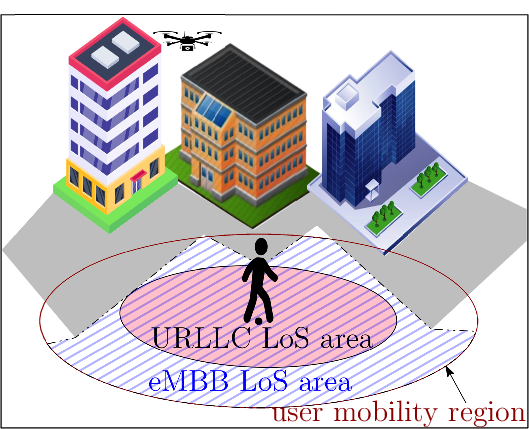}
    \caption{}
    \label{fig:environment}
\end{subfigure}
\begin{subfigure}{0.15\textwidth}
    \includegraphics[width=\linewidth]{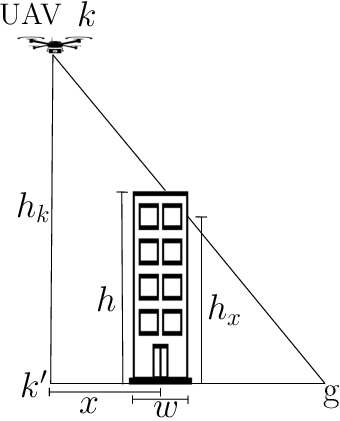}
    \caption{}
    \label{fig:blockage}
\end{subfigure}
\caption{(a) LoS for URLLC and eMBB (b) LoS Blockage}
\vspace{-2mm}
\end{figure}
\vspace{-2mm}
\subsection{Blockage Model}
\noindent Here we restrict ourselves to consider only building blockages whose centers are distributed according to the homogeneous PPP $\Phi_B$ with density $\lambda_B$. The length $l$ and width $w$ of each building are taken from uniform $\mathcal{U}\;(L_{\rm min},L_{\rm max})$, where $L_{\rm min}$ \& $L_{\rm max}$ are the dimension bounds of the buildings. The height of each building is taken from the Rayleigh distribution with density function 
$f(h)=\frac{h}{\gamma^2}e^{\frac{-h^2}{2\gamma^2}}$, where $\gamma$ is an environment dependent parameter \cite{al2014optimal}. The building blockages can be visualized as cuboids.

\subsection{Communication Model}
\noindent In mmWave communication, penetration loss is very high with physical obstacles. So, we consider that a user will be served by UAV $k$ only if it has LoS with the UAV and falls within its communication range $R_k$.
Note that given the highly directional and spatially compact characteristics of the beamformed signal, multipath effects such as scattering and reflection can be reasonably disregarded to simplify the channel modeling.
Consider a particular user-UAV ($g-k$) link. As per the $73$ GHz model, the path-loss function (in dB) is,
\begin{equation}
    PL_{g,k}= \alpha+10 \beta\log_{10}(d_{g,k}),
\end{equation}
where $d_{g,k}$ is the Euclidean distance between the user $g$ and the UAV $k$, $\alpha$ and $\beta$ are environment dependent parameters \cite{channelmodel}. Depending on the path loss model, the received power (in dBm) at $g$ from UAV $k$ becomes,
\begin{equation}
    P_{g,k}= P_k+G_k+G_g-PL_{g,k},
\end{equation}
where $P_k$ is the transmit power of UAV $k$, $G_k$ and $G_g$ are the antenna gain at the UAV end and the user end respectively.
Hence, the throughput obtained at user $g$ from UAV $k$ is,
\begin{equation}\label{thrput}
    T_k(g)= B_w\log_2\left(1+\frac{P_{g,k}|h_{g,k}|^2}{\sigma_0}\right),
\end{equation}
where $B_w$ is the channel bandwidth, $|h_{g,k}|$ follows a Rician distribution following LoS characteristics \cite{lakshmirayleigh}, and $\sigma_0$ denotes the variance of the additive white Gaussian noise.

\subsection{User Traffic}
\noindent For eMBB users, maximizing the average data rate is of primary importance. Since eMBB traffic can tolerate slight delays, it is sufficient to consider the average area over which a LoS connection is maintained within the user mobility region. Accordingly, we define the \textit{eMBB LoS area} as the area of the expected region over which a user maintains uninterrupted LoS connectivity within the user mobility region.
In contrast, for URLLC communication scenarios, maintaining a stable LoS connection is imperative, regardless of user mobility. To ensure uninterrupted LoS connectivity in all possible movement directions inside the user mobility region, we define the \textit{URLLC LoS radius} as the radius of the largest circle within which the user consistently maintains a seamless LoS link to the UAV. Note that due to the small packet size of URLLC users, as long as the user maintains LoS connectivity with the UAV, the throughput requirement of the user is satisfied. Thus, the eMBB and URLLC user will be associated to the UAV that maximizes the average throughput based on the eMBB LoS area and the URLLC LoS radius, respectively.

\section{Analytical Approach}\label{analytical}

\noindent Let the link between the ground user $g=(x_g,y_g,h_g)$ and the UAV $k=(x_k,y_k,h_k)$ be $gk$. For simplicity of calculation, we take the height of the ground user as $h_g=0$. Let the 2D projection of the link $gk$ (as in Fig. \ref{fig:blockage}) be $gk'$, where $g=(x_g,y_g,0)$ and $k'=(x_k,y_k,0)$. Let $|gk'|=a_k$ be the length of the link $gk'$. Clearly, a building with width $w$ and height $h$ blocks the link $gk$ only if $h>h_x$, where $h_x$
\begin{equation}
    \!\!\!= \frac{[a_k-(x+\frac{w}{2})]h_{k}+(x+\frac{w}{2})h_g}{a_k}\overset{(a)}{=} \frac{[a_k-(x+\frac{w}{2})]h_{k}}{a_k},\!
\end{equation}
where $(a)$ follows from the fact that $h_g=0$. Here $x$ is the 2D distance from $k'$ to the center of the building blocking the link.
Given a building of width $w$ that blocks $gk'$, the probability that it also blocks $gk$ is
\begin{align}
    \xi^k&= \frac{1}{a_k}\int_0^{a_k}\mathbb{P}[h>h_x]dx\nonumber\\
    &= \frac{1}{a_k}\int_0^{a_k}\left[1-\int_0^ \frac{\left[a_k-\left(x+\frac{w}{2}\right)\right]h_{k}}{a_k}f(h)dh\right]dx\nonumber \\
    &= 1-\frac{1}{a_k}\int_0^{a_k}\int_0^{\left(1-\frac{x+\frac{w}{2}}{a_k}\right)h_{k}}\frac{h}{\gamma^2}e^{\frac{-h^2}{2\gamma^2}}dh dx\nonumber \\
    \!\!\!\!&= \sqrt{\frac{\pi}{2}}\frac{\gamma}{h_{k}}\left[erf\!\left(\frac{w h_{k}}{2 \sqrt{2}a_k\gamma}\right)\!-erf\!\left(\frac{(w-2a_k)h_{k}}{2 \sqrt{2}a_k\gamma}\right)\right]
\end{align}
where $erf(.)$ is the error function defined by, $erf(x)=\frac{2}{\sqrt{\pi}}\int_0^xe^{-t^2}dt$. Also, clearly $\xi^k<1.$

 From \cite{bai2014analysis}, without considering the building heights, the probability that a link of length $a_k$ is in LoS is  $e^{-(\zeta a_k+\tau)}$, where $\zeta= 2\lambda_B\frac{\mathbb{E}[w]+\mathbb{E}[l]}{\pi}$ and $\tau = \lambda_B\mathbb{E}[w]\mathbb{E}[l]$.  So, considering building heights, the probability $p(a_k)$
that a link of length $a_k$ is in LoS is $e^{-\left(\frac{\zeta}{\xi^k} a_k+\frac{\tau}{\xi^k}\right)}$. This expression is obtained by replacing $\lambda_B$ with $\frac{\lambda_B}{\xi^k}$ in $\zeta$ and $\tau$, as incorporating the building heights introduces a constant scaling factor $\xi^k$.

In the next two subsections, assuming that the blockages affect the links independently, we compute the eMBB LoS area and URLLC LoS radius using the probability $p(a_k)$.
\subsection{eMBB LoS Area Calculation}
\noindent We calculate the eMBB LoS area of the user $g$ for each UAV $k$ as stated in the following lemma.
\begin{lemma}
    The eMBB LoS area $\Delta_k$ is $\min\{\frac{2\pi e^{-\frac{\tau}{\xi^k}}}{\left(\frac{\zeta}{\xi^k}\right)^2},\pi R^2_g\}$.
\end{lemma}

\begin{proof}
    Let $B_{xy} $ denote the number of blockages on a link $xy$, where $x,y \in \mathbb{R}^2$. Let $A(L_g)$ be the area of 
    the LoS region of $g$ specified by $L_g= \{r\in\mathbb{R}^2: B_{gr}= 0\}$.
    Alternatively, $B_{xy}$ can be expressed in polar coordinate system as $B_{l\phi}$, where $l$ is the length and $\phi$ is the angle of the link with the positive $x$-axis.
    Let $P$ denote the event that the user is not covered by any blockage and $P^c$ be its complement event.
   Then $\mathbb{P}\{P\}= p(0)=e^{\frac{-\tau}{\xi^k}}$ \cite{bai2014analysis}. Now, the distance to the nearest blockage from $g$ along $\phi$ is $D_\phi= sup\{t\in\mathbb{R^+}: B_{t\phi}= 0\}$. As blockages affect links independently, we have $\mathbb{P}\{B_{a_k\phi}=0\}=p(a_k)=e^{-\left(\frac{\zeta}{\xi^k}a_k+\frac{\tau}{\xi^k}\right)}$.
    Now we get,
    \begin{align}
        \mathbb{P}\{D_\phi>a_k|P\}
        =\frac{\mathbb{P}\{B_{a_k\phi}=0\}}{\mathbb{P}\{P\}}
        =\frac{p(a_k)}{p(0)}
        = e^{-\frac{\zeta}{\xi^k}a_k}.
    \end{align}
    Now, the average LoS area is,
      $\Delta'_k =\mathbb{E}[A(L_g)]$
      \begin{align}
      &= \mathbb{P}\{P\}\times\mathbb{E}[A(L_g)|P]+(1-\mathbb{P}\{P\})\times \mathbb{E}[A(L_g)|P^c]\nonumber \\
        &\overset{(a)}{=} \mathbb{P}\{P\}\times\mathbb{E}\left[\int_0^{2\pi}\frac{D^2_\phi}{2}d\phi|P\right]\nonumber \\
        &= 2\pi e^{-\frac{\tau}{\xi^k}}\int_0^{\infty}\frac{x^2}{2}\frac{\zeta}{\xi^k}e^{-\frac{\zeta}{\xi^k}x}dx
        = \frac{2\pi e^{-\frac{\tau}{\xi^k}}}{\left(\frac{\zeta}{\xi^k}\right)^2},
    \end{align}
    where $(a)$ follows from the fact that if the location of user $g$ is covered by any blockage, then $A(L_g)= 0$. 

Since we consider the communication only in the $\Delta t$ duration, the actual eMBB LoS area $\Delta_k$ is given by, 
\begin{equation}
    \Delta_k = \min\{\Delta'_k, \pi R^2_g\}
    = \min\{\frac{2\pi e^{-\frac{\tau}{\xi^k}}}{\left(\frac{\zeta}{\xi^k}\right)^2},\pi R^2_g\},
\end{equation}
where $R_g=v_g\Delta t$, $v_g$ being the speed of the user during $\Delta t.$
\end{proof}

\subsection{URLLC LoS Radius Calculation}
\noindent We calculate the URLLC LoS radius of the user $g$ for each UAV $k$ as stated in the following lemma.
\begin{lemma}
    The URLLC LoS radius $R^k_{\rm URLLC}$ is $\min\{\frac{\sqrt{\xi^k}}{\sqrt{2\pi\lambda_B}}, R_g\}.$
\end{lemma}
\begin{proof}

For a homogeneous PPP $\Phi_B^k$ of intensity $\lambda_B^k=\frac{\lambda_B}{\xi^k}$ on $\mathbb{R}^2$, the distribution of total building points inside the disk of radius $r$ is Poisson with mean $\lambda_B^k\pi r^2$.
In stochastic geometry framework, the contact distance at a location $g$ of a PPP $\Phi_B^k$ is defined by $||g-\Phi_B^k||=\inf\{||g-z||: z \text{ is generated from PPP }\Phi_B^k\}$ \cite{haenggi2013stochastic}.
Hence, the contact distribution function is,
\begin{equation}
    \mathbb{F}^g(r)= \mathbb{P}\left(||g-\Phi_B^k||\leq r\right)=1-e^{-\lambda_B^k\pi r^2}.
\end{equation}
Then the density function becomes,\vspace{-1mm}
\begin{equation}
    f(r)= 2\pi\lambda_B^k r e^{-\lambda_B^k\pi r^2}.
    \vspace{-1mm}
\end{equation}
The most probable contact distance, which is the mode of the distribution is obtained by solving
\vspace{-1mm}
\begin{equation}
    \!\!\!\!\!\!f'(r)=0
    \implies 2\pi\lambda_B^k e^{-\lambda_B^k\pi r^2}\left(1-2\pi\lambda_B^k r^2\right)= 0.
    \vspace{-1mm}
\end{equation}
\vspace{-1mm}
\begin{equation}
\!\!\!\!\!\!\!\!\!\!\!\!\!\!\!\!\!\!\!\!\!\!\!\!\!\!\!\!\!\!\!\!\!\!\!\!\!\!\!\text {Solving, we get,}\;\; r= \frac{1}{\sqrt{2\pi\lambda_B^k}}=\frac{\sqrt{\xi^k}}{\sqrt{2\pi\lambda_B}}.
\end{equation}
Since we consider communication only in the $\Delta t$ duration, the user cannot go beyond its mobility region, and hence the actual URLLC LoS radius is, \begin{equation}
    R^k_{\rm URLLC}= \min\{r, R_g\}=\min\{\frac{\sqrt{\xi^k}}{\sqrt{2\pi\lambda_B}},R_g\}.\nonumber\quad\quad\quad\quad\qedsymbol\qedhere
    \end{equation}
\end{proof}
\section{Shadow Polygon Approach}\label{spa}

\noindent The results in previous sections are derived based on the independent link assumption. However, this may not hold in practice. Note that if two links originating from the same user share a common path, the larger link have no fewer blockages than the shorter one, implying that the links are correlated. To overcome this limitation, here we present a shadow polygon approach to calculate the accurate eMBB LoS area and URLLC LoS radius geometrically where the link independence assumption is not considered.
In this method, the UAV $k$ will use the map of the environment under consideration as input for some preprocessing calculations at the start of $\Delta t$ time. The system can easily obtain buildings' horizontal coordinates via google maps and also they can obtain building heights by comparing the elevation of the building with the ground.
\subsubsection{\bf Preprocessing}
\noindent As mentioned in the system model, the building blockages are taken as cuboids. Let the coordinates of the vertices of $l^{th}$ building be $\{\left(x^l_i,y^l_i,h^l_i\right):i=1,2,\dots,8\}$. Let $\{(x_i,y_i,h_i)\}_{i=5}^8$  be the building vertices lying on the ground (for simplicity, we omit the $l$ for now). That is, $h_i=0$ for $i=5,\dots,8$. So the equation of the line connecting the building vertices of non-zero height, i.e., $\{(x_i,y_i,h_i)\}_{i=1}^4$  and the UAV coordinates $(x_k,y_k,h_k)$ be
\begin{equation}\label{line}
    \frac{x-x_i}{x_i-x_k}= \frac{y-y_i}{y_i-y_k}= \frac{h-h_i}{h_i-h_k}.
\end{equation}

Now if we intersect the line \eqref{line} with the plane $h=0$ the user currently resides on, we get,
\begin{align}
    x= x_i-h_i\frac{x_i-x_k}{h_i-h_k}, \;
    y= y_i-h_i\frac{y_i-y_k}{h_i-h_k},\;
    h= 0. 
\end{align}
Using the above equations, we get four points $\{(x'_i,y'_i,0)\}_{i=1}^4$ which are the projection  of the vertices $\{(x_i,y_i,h_i)\}_{i=1}^4$ with non-zero heights, on the ground with respect to the UAV $k$. Also, for the ground vertices of the building, we get $(x'_i,y'_i,0)=(x_i,y_i,0)\; \forall \;i=5,\dots,8$. Thus we have $8$ points on 2D plane $\{(x'_i,y'_i)\}_{i=1}^8$ with respect to each building. We can easily get the following lemma.
\vspace{-2mm}
\begin{lemma}
The shadow region of the cuboid building with respect to an UAV is the convex hull of the projections of all vertices of the cuboid with respect to the UAV.
\end{lemma}
\vspace{-4mm}
\begin{proof}
    Let us prove this lemma by the method of contradiction. Let the shadow region $\mathcal{S}$ not be the convex hull of the projected points. This implies that there exists at least a point $(x'_p,y'_p) \notin\mathcal{S}$ which is on the line segment joining $(x'_i,y'_i)$ and $(x'_j,y'_j)\in\mathcal{S}$ for some $i \neq j$. Then there must also exist two points $(x_i,y_i,h_i)$ and $(x_j,y_j,h_j)$ on the cuboid building such that $(x_p,y_p,h_p)$ lies outside the cuboid building. This implies that the cuboid is not convex, a contradiction.
\end{proof}
\vspace{-2mm}
There exist well-known $\mathcal{O}(n\log{}n)$ time algorithms to calculate convex hull, given $n$ points in 2D plane \cite{cormen199033}. Depending on the position of the UAV and the cuboid, the shadow will be a $p$-gon with $p=4,5,6$. Then, we will compute the union of all these shadow convex hulls. There exist $\mathcal{O}(n\log{}n+k)$ time algorithm to calculate the union of two polygons using line sweep method where $n$ is the number of total vertices and $k$ is the number of points they intersect \cite{O’Rourke_1998}. Doing this operation for all the polygons, we obtain non-intersecting shadow polygons (they may not be convex). Essentially, in this preprocessing step, we calculate the shadow polygons corresponding to each UAV.

\subsubsection{\bf Computation of eMBB LoS Area}
\noindent At the beginning of each $\Delta t$ interval, the UAV $k$ will obtain 
the user coordinates $g(x_g,y_g)$ (we omit the height as $h_g=0$) and compute the eMBB LoS area $\Delta_k$ accordingly via Algorithm \ref{alg:eMBB}. Here, basically we want to calculate the area of the region that the user can "see" given the environment filled with simple and bounded non-convex polygonal walls. Now, we can consider a polygonal approximation of user's circular mobility region as the polygon and the shadow polygons as holes of that polygon.
Then our problem reduces to: \textit{given a polygon with simple bounded holes and a point inside the polygon, what is the visibility polygon of that point?} This visibility polygon can be calculated using an $\mathcal{O}(n \log h)$ time algorithm \cite{Ghosh_2007} based on angular sweep method, where $h$ is the number of holes and $n$ is the number of total vertices of the polygon. 
After getting the visibility polygon, we can obtain its area by using the well-known Gauss's area formula for irregular polygon.
\begin{algorithm}
\caption{eMBB LoS Area Calculation}\label{alg:eMBB} 
\KwIn{Shadow polygons obtained through preprocessing, user coordinates $g(x_g,y_g)$} 
\KwOut{eMBB LoS area $\Delta_k$}
\For{Each UAV $k$}{
    Calculate the visibility polygon $V_k(g)$ as in \cite{Ghosh_2007};\\
    Find the area $\Delta_k$ of $V_k(g)$ by Gauss's formula.\\
    }
\end{algorithm}

\textbf{\textbullet \- Complexity Analysis:}
In RAM model, the $V_k(g)$ takes $\mathcal{O}(n \log h)$ time. The area $\Delta_k$ calculation by Gauss's formula  takes unit time. Hence, the worst case time complexity of the algorithm is $\mathcal{O}(mn\log h)$, where $m$ is the number of UAVs.
\subsubsection{\bf Computation of URLLC LoS Radius}
\noindent At the beginning of the $\Delta t$ time slot, each UAV can obtain user coordinates $g(x_g,y_g)$ and then calculate the URLLC LoS radius using Algorithm \ref{alg:urllc}. By calculating the minimum of the shortest distances $d^j_s$ from $g$ to all line segments $l^j_s$ of each shadow polygon $j$, the algorithm determines the minimum distance $d^j$ from $g$ to each shadow polygon $j$. The well-known vector projection approach is used to compute $d^{j}_{s}$. Lastly, the URLLC LoS radius $R^k_{\rm URLLC}$ is computed as $R^k_{\rm URLLC}= \min\{\displaystyle\min_{j}\;d^j, R_g\}$.
\begin{algorithm}
\caption{URLLC LoS Radius Calculation}\label{alg:urllc} 
\KwIn{Shadow polygons obtained through preprocessing, user coordinates $g(x_g,y_g)$} 
\KwOut{URLLC LoS radius $R^k_{\rm URLLC}$}
\For{Each UAV $k$}{
\For{Each shadow polygon $j$}{
    \For{Each line segments $l^j_s$ }{
         Calculate the distance $d^{j}_{s}$ from $g$ to $l^j_s$ using vector projection approach;\\
    }
    $d^j= 
    \displaystyle\min_{s}\; d^{j}_{s}$\\
}
$R^k_{\rm URLLC}= \min\{\displaystyle\min_{j}\;d^j, R_g\}.$
}
\end{algorithm}

\textbf{\textbullet \- Complexity Analysis:} 
In RAM model, computation of $d^{j}_{s}$ using vector projection method takes unit time. The first for loop runs for $m$ times, where $m$ is the number of UAVs. The second for loop runs $q$ times, where $q$ is the number of shadow polygons. If $c$ is maximum number of edges in any shadow polygon, then the third for loop runs $c$ times. Hence, the worst case time complexity is $\mathcal{O}(cmq)$.

\subsection{UAV Association Policy}
\noindent Now we associate the user $g$ to that UAV which corresponds to maximum average throughput and maximum URLLC LoS radius, for eMBB and URLLC user respectively.
\subsubsection{\bf Association for eMBB User}
\noindent To be able to communicate with UAV $k$, the user $g$ must reside within the circle $C_k$ of radius $R'_k=\sqrt{R^2_k- {h^2_k}}$ centering $(x_k,y_k,0)$, where $R_k$ is the communication range of UAV $k$ and $h_k$ is its height. Let $P_k$ be the polygonal approximation of the circle $C_k$. Note that the user $g$ must also be in its visibility polygon $V_k(g)$ computed by Algorithm \ref{alg:eMBB}. Let $I_k(g)= P_k \cap V_k(g)$ and $A_k(g)$ be the area of $I_k(g)$. Note that $I_k(g)$ can be obtained by using line sweep based $\mathcal{O}(r\log r+t)$ algorithm \cite{O’Rourke_1998}, where $r$ is the number of total vertices of the polygons $P_k$ and $V_k(g)$, and $t$ is the number of points they intersect. Now we triangulate $I_k(g)$ using well-known $\mathcal{O}(p\log p)$ algorithm ($p$ being the number of vertices of $I_k(g)$) \cite{deBerg2008} and obtain $A_i$ such that $A_k(g)=\sum_{i=1}^N A_i$, where $A_i$ denotes the area of triangle $i$ and $N$ is number of triangles. 

Assuming that a user does not get any data rate if it is not having the LoS with the UAV, the expected throughput obtained by user $g$ from UAV $k$ can be computed as $\mathbb{E}_k(g)=\frac{A_k(g)}{\pi R_g^2} T_k(g)+ (1-\frac{A_k(g)}{\pi R_g^2}) \times 0=\frac{A_k(g)}{\pi R_g^2} T_k(g)$, where $T_k(g)$ is the throughput obtained by user $g$ from UAV $k$. Now we can compute $T_k(g)= \frac{\sum_{i=1}^N (T_k(c_i)\times A_i)}{\sum_{i=1}^N A_i}$, where $c_i$ is the centroid of triangle $i$ and $T_k(c_i)$ is the throughput at $c_i$ from UAV $k$, as computed using equation \eqref{thrput}. Hence the expected throughput becomes, $\mathbb{E}_k(g)=\frac{\sum_{i=1}^N (T_k(c_i)\times A_i)}{\pi R^2_g}$. Finally, the user $g$ will be associated with the UAV $k^*=\displaystyle\arg \max_{k}\mathbb{E}_k(g)$ that provides the maximum expected throughput. 

\subsubsection{\bf Association for URLLC User}
\noindent To be able to communicate with UAV $k$, the user $g$ must be inside the circle $C_k$ of radius $R'_k$ and also within the circle of radius $R^k_{\rm URLLC}$. In other words, $R^k_{\rm URLLC}$ must be completely inside $C_k$. That is, $R^k_{\rm URLLC} + d'_{g,k} \leq R'_{k}$, where $d'_{g,k}$ is the 2D distance between $(x_g,y_g)$ and $(x_k,y_k)$. Finally, the user $g$ will be associated with the UAV $k^*= \displaystyle\arg \max_{k}\{R^k_{\rm URLLC}: R^k_{\rm URLLC} + d'_{g,k} \leq R'_{k}\}$ that provides the maximum URLLC LoS radius.

\section{Simulation Results}\label{simulation}
\begin{figure*}
\centering
\begin{subfigure}[b]{0.24\textwidth}
\centering
\includegraphics[width=\linewidth]{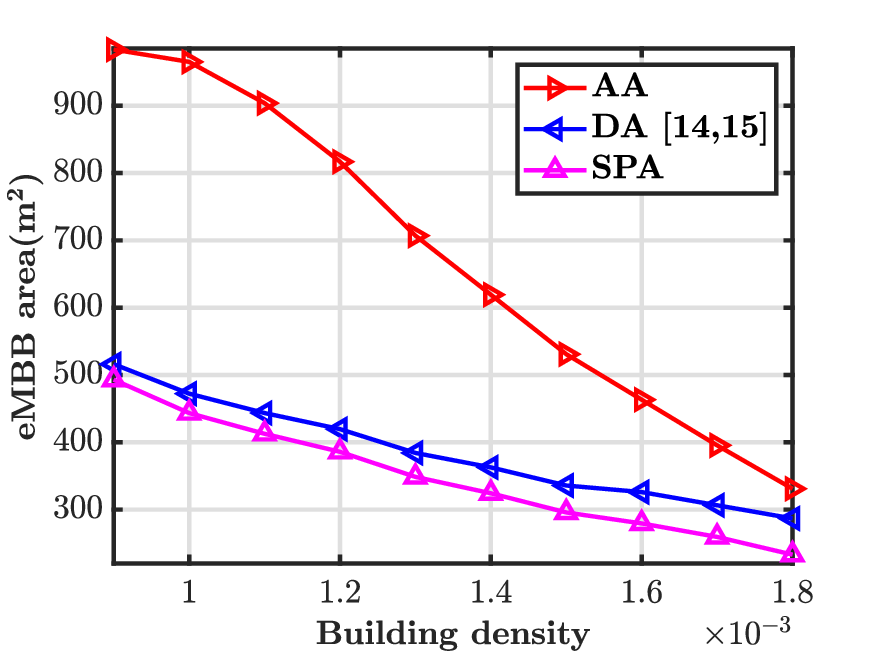}
\vspace{-2mm}
\caption{}
\label{fig:embb area}
\vspace{-2mm}
\end{subfigure}
\begin{subfigure}[b]{0.24\textwidth}   
\includegraphics[width=\linewidth]{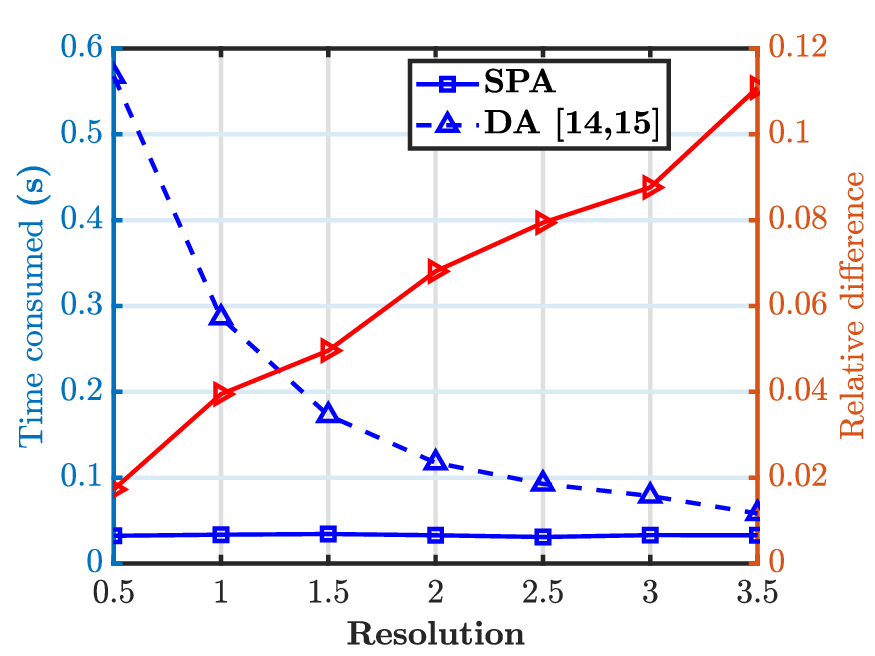}
\vspace{-2mm}
\caption{}
\label{fig:embb time}
\vspace{-2mm}
\end{subfigure}
\begin{subfigure}[b]{0.24\textwidth}   
\includegraphics[width=\linewidth]{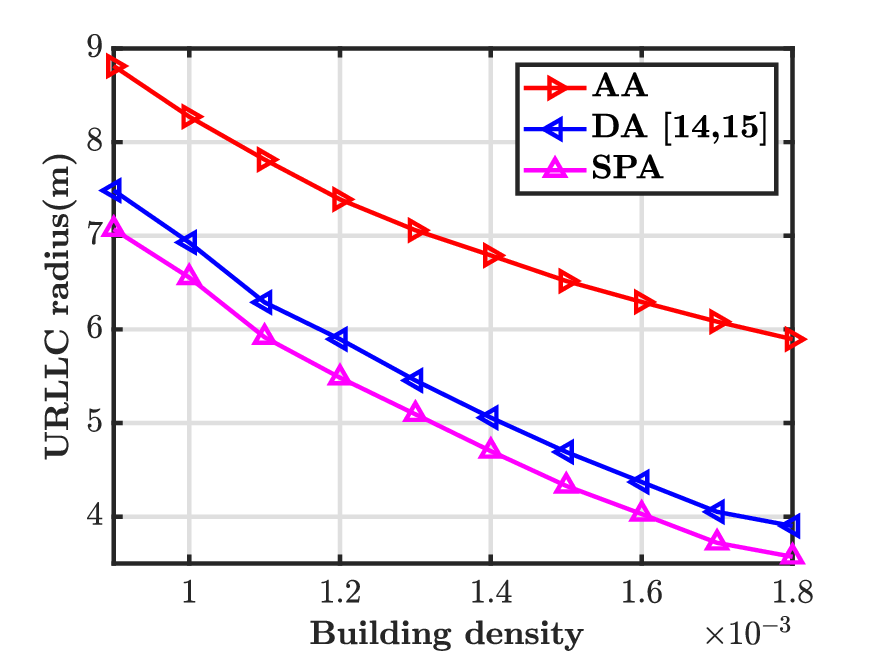}
\vspace{-2mm}
\caption{}
\label{fig:URLLC Radius}
\vspace{-2mm}
\end{subfigure}
\begin{subfigure}[b]{0.24\textwidth}
\centering
\includegraphics[width=\linewidth]{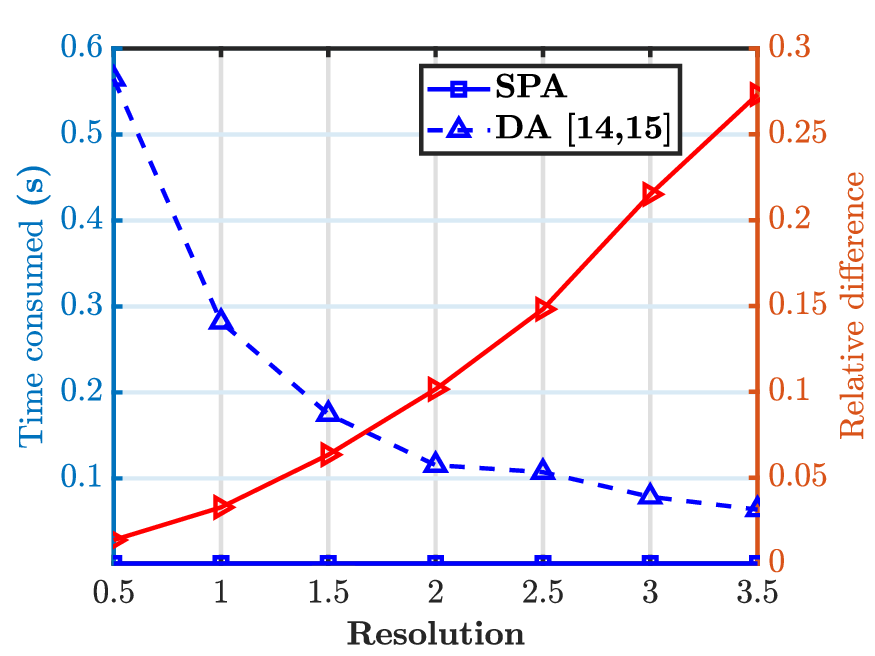}
\vspace{-2mm}
\caption{}
\label{fig:urllc time comparison}
\vspace{-2mm}
\end{subfigure}
\caption{(a) Building density vs eMBB area for $a=0.5$, (b) Resolution vs time and relative difference of eMBB area, 
(c) Building density vs URLLC radius for $a=0.5$, (d) Resolution vs time and relative difference of URLLC radius}
\vspace{-4mm}
\end{figure*}
\noindent Here, we compare our shadow polygon approach (SPA) with analytical approach (AA) as well as a discretization approach (DA) considered in \cite{sabzehali20213d, li2022geometric}. In the latter case, the service area is discretized into $a\times a$ squares, where the LoS status of the square with respect to the UAV is determined based on that of its center. The parameter $a$ can be visualized as the resolution parameter, where a smaller value indicates higher resolution. The simulation parameters are given in Table \ref{tab:simulation parameters}.
\begin{table}[h!]
\centering
\begin{tabular}{|c|c|}
\hline
\textbf{Parameter} & \textbf{Value}\\
\hline
Communication Region & $400m\times400m$\\
\hline
Slot Duration ($\Delta t$)  & 5s\\
\hline
User Velocity ($v_g$) & [2,4]m/s\\ 
\hline
UAV communication range & 250 m\\
\hline
Building length \& width & [10,20]m\\
\hline
Height parameter ($\gamma$) & 7.63\\
\hline
Carrier frequency & 73 GHz \cite{channelmodel}\\
\hline
Rician fading factor & 2 \cite{sau2024drams}\\
\hline
LoS link parameters & $\alpha=69.8$ , $\beta=2$ \cite{channelmodel}\\
\hline
Transmit Power & 30 dBm \cite{sau2024drams}\\
\hline
Antenna gain & 24.5 dBi \cite{channelmodel}\\
\hline
\end{tabular}
\caption{Simulation Parameters}
\label{tab:simulation parameters}
\end{table}

Fig. \ref{fig:embb area} shows that as building density increases, the average eMBB area decreases following a similar trend for all three approaches. Due to the link independence assumption, the value obtained by AA differs from that of SPA. Note that SPA is supposed to be more accurate as it computes the eMBB area without the link independent assumption. The difference in the values obtained by DA and SPA depends on the resolution. The difference becomes smaller as $a$ decreases. Use of higher values of $a$ requires less computing time but leads to higher difference as shown in Fig. \ref{fig:embb time}. We observe a similar phenomenon in Fig. \ref{fig:URLLC Radius} for URLLC radius computed by the three approaches. Here also the difference between the values obtained by DA and SPA increases with $a$ while the computation time of the former decreases as shown in Fig. \ref{fig:urllc time comparison}. 

\begin{figure}
\centering
\begin{subfigure}[b]{0.25\textwidth}
\centering
\includegraphics[width=\linewidth]{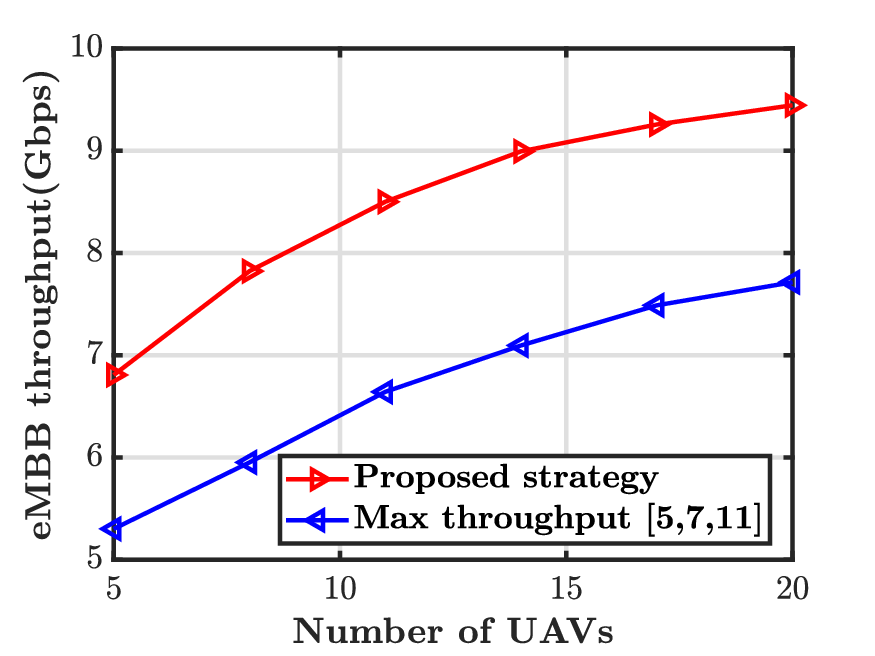}
\caption{}
\label{fig:comparison embb throughput}
\vspace{-2mm}
\end{subfigure}
\hspace{-0.5cm}
\begin{subfigure}[b]{0.25\textwidth}   
\includegraphics[width=\linewidth]{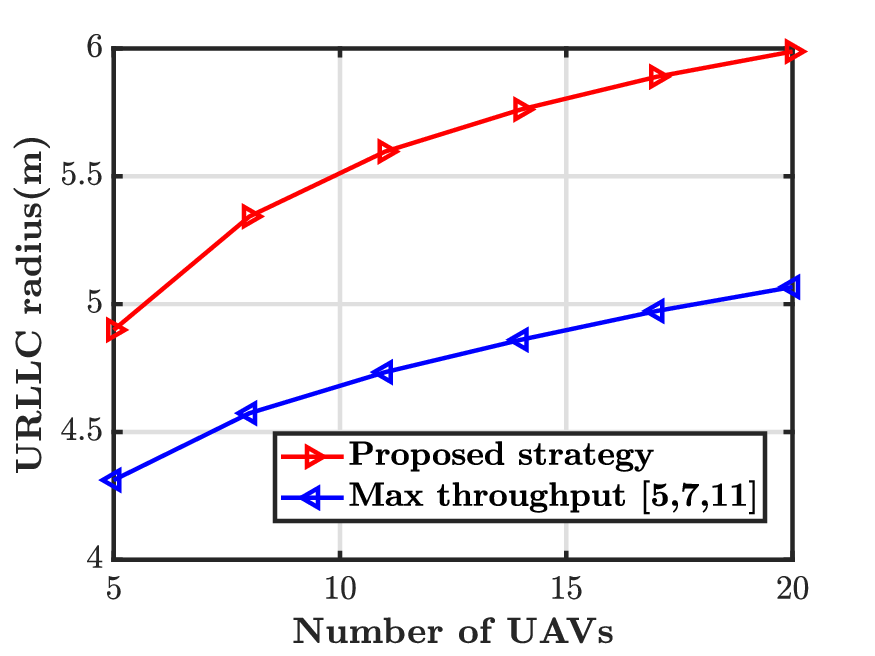}
\caption{}
\label{fig:comparison URLLC Radius}
\vspace{-2mm}
\end{subfigure}
\vspace{-4mm}
\caption{UAVs vs (a) eMBB throughput (b) URLLC radius}
\label{fig: comparison }
\vspace{-5mm}
\end{figure}

As explained earlier, in studies \cite{wang2020trajectory,mahmood2023joint,zhang2024joint} the user is associated to that UAV which gives maximum throughput at that point. But we can clearly see in Fig. \ref{fig:comparison embb throughput}, for building density $\lambda_B=1.2\times10^{-3}$, the average throughput across the whole time interval is larger if we associate the UAV in terms of maximum average throughput based on eMBB LoS area. Also, it is quite intuitive to see that the average throughput increases with the number of UAVs. Similarly, in case of URLLC users, Fig. \ref{fig:comparison URLLC Radius} shows the superiority of our strategy compared to maximum throughput based association policies by showing that our association policy giving longer URLLC radius to a user than the maximum throughput based strategy.
\vspace{-3mm}
\section{Conclusion}\label{conclusion}
\vspace{-1mm}
\noindent In this work, we investigated the impact of blockage density on the UAV association with dynamic user for different traffic types. We calculated the eMBB area and URLLC radius both analytically and using geometric shadow polygon approach. Then we associated the user with UAV with largest average throughput and longest URLLC radius for eMBB and URLLC user respectively. The numerical results demonstrated the superiority of our approach over some existing approaches. 
 \vspace{-2mm}
\bibliographystyle{IEEEtran}
\bibliography{main}

\begin{thebibliography}{10}
\providecommand{\url}[1]{#1}
\csname url@samestyle\endcsname
\providecommand{\newblock}{\relax}
\providecommand{\bibinfo}[2]{#2}
\providecommand{\BIBentrySTDinterwordspacing}{\spaceskip=0pt\relax}
\providecommand{\BIBentryALTinterwordstretchfactor}{4}
\providecommand{\BIBentryALTinterwordspacing}{\spaceskip=\fontdimen2\font plus
\BIBentryALTinterwordstretchfactor\fontdimen3\font minus \fontdimen4\font\relax}
\providecommand{\BIBforeignlanguage}[2]{{%
\expandafter\ifx\csname l@#1\endcsname\relax
\typeout{** WARNING: IEEEtran.bst: No hyphenation pattern has been}%
\typeout{** loaded for the language `#1'. Using the pattern for}%
\typeout{** the default language instead.}%
\else
\language=\csname l@#1\endcsname
\fi
#2}}
\providecommand{\BIBdecl}{\relax}
\BIBdecl

\bibitem{ericsson}
{E}ricsson mobility report, Nov. 2024, [Online]. Available: https://www.ericsson.com/en/reports-and-papers/mobility-report/reports/november-2024.

\bibitem{navarro2020survey}
J.~Navarro-Ortiz, P.~Romero-Diaz, S.~Sendra, P.~Ameigeiras, J.~J. Ramos-Munoz, and J.~M. Lopez-Soler, ``A survey on 5{G} usage scenarios and traffic models,'' \emph{IEEE Commun. Surv. Tutorials}, vol.~22, no.~2, pp. 905--929, Feb. 2020.

\bibitem{bai2014analysis}
T.~Bai, R.~Vaze, and R.~W. Heath, ``Analysis of blockage effects on urban cellular networks,'' \emph{IEEE Trans. Wireless Commun.}, vol.~13, no.~9, pp. 5070--5083, Oct. 2014.

\bibitem{li2018uav}
B.~Li, Z.~Fei, and Y.~Zhang, ``{UAV} communications for 5{G} and beyond: {R}ecent advances and future trends,'' \emph{IEEE Internet Things J.}, vol.~6, no.~2, pp. 2241--2263, June 2018.

\bibitem{zhang2024joint}
R.~Zhang, Y.~Zhang, R.~Tang, H.~Zhao, Q.~Xiao, and C.~Wang, ``A joint {UAV} trajectory, user association, and beamforming design strategy for multi-{UAV} assisted {ISAC} systems,'' \emph{IEEE Internet Things J.}, 2024.

\bibitem{singh2022user}
S.~K. Singh, V.~S. Borkar, and G.~S. Kasbekar, ``User association in dense mmwave networks as restless bandits,'' \emph{IEEE Trans. Veh. Technol.}, vol.~71, no.~7, pp. 7919--7929, July 2022.

\bibitem{mahmood2023joint}
A.~Mahmood, T.~X. Vu, S.~Chatzinotas, and B.~Ottersten, ``Joint optimization of {3D} placement and radio resource allocation for per-{UAV} sum rate maximization,'' \emph{IEEE Trans. Veh. Technol.}, vol.~72, no.~10, pp. 13\,094--13\,105, Oct. 2023.

\bibitem{van2022urllc}
D.~Van~Huynh, V.-D. Nguyen, S.~R. Khosravirad, V.~Sharma, O.~A. Dobre, H.~Shin, and T.~Q. Duong, ``{URLLC} edge networks with joint optimal user association, task offloading and resource allocation: {A} digital twin approach,'' \emph{IEEE Trans. Commun.}, vol.~70, no.~11, pp. 7669--7682, Nov. 2022.

\bibitem{xi2020network}
X.~Xi, X.~Cao, P.~Yang, J.~Chen, T.~Q. Quek, and D.~Wu, ``Network resource allocation for e{MBB} payload and {URLLC} control information communication multiplexing in a multi-{UAV} relay network,'' \emph{IEEE Trans. Commun.}, vol.~69, no.~3, pp. 1802--1817, Mar. 2020.

\bibitem{tian2023demand}
M.~Tian, C.~Li, Y.~Hui, N.~Cheng, W.~Yue, Y.~Fu, and Z.~Han, ``On-{D}emand {M}ultiplexing of e{MBB/URLLC} {T}raffic in a {M}ulti-{UAV} {R}elay {N}etwork,'' \emph{IEEE Trans. Intell. Transp. Syst.}, vol.~25, no.~6, pp. 6035--6048, June 2023.

\bibitem{wang2020trajectory}
Y.-S. Wang, Y.-W.~P. Hong, and W.-T. Chen, ``Trajectory learning, clustering, and user association for dynamically connectable {UAV} base stations,'' \emph{IEEE Trans. Green Commun. Networking}, vol.~4, no.~4, pp. 1091--1105, Apr. 2020.

\bibitem{cheng2024learning}
Z.~Cheng, M.~Liwang, N.~Chen, L.~Huang, N.~Guizani, and X.~Du, ``Learning-based user association and dynamic resource allocation in multi-connectivity enabled unmanned aerial vehicle networks,'' \emph{Digital Commun. Networks}, vol.~10, no.~1, pp. 53--62, Jan. 2024.

\bibitem{guan2021user}
X.~Guan, Y.~Huang, C.~Dong, and Q.~Wu, ``User association and power allocation for {UAV}-assisted networks: {A} distributed reinforcement learning approach,'' \emph{China Commun.}, vol.~17, no.~12, pp. 110--122, 2021.

\bibitem{sabzehali20213d}
J.~Sabzehali, V.~K. Shah, H.~S. Dhillon, and J.~H. Reed, ``3{D} placement and orientation of mm{W}ave-based {UAV}s for guaranteed {L}o{S} coverage,'' \emph{IEEE Wireless Commun. Lett.}, vol.~10, no.~8, pp. 1662--1666, Oct. 2021.

\bibitem{li2022geometric}
F.~Li, C.~He, X.~Li, J.~Peng, and K.~Yang, ``Geometric analysis-based 3{D} anti-block {UAV} deployment for mm{W}ave communications,'' \emph{IEEE Commun. Lett.}, vol.~26, no.~11, pp. 2799--2803, Nov. 2022.

\bibitem{al2014optimal}
A.~Al-Hourani, S.~Kandeepan, and S.~Lardner, ``Optimal {LAP} altitude for maximum coverage,'' \emph{IEEE Wireless Commun. Lett.}, vol.~3, no.~6, pp. 569--572, June. 2014.

\bibitem{channelmodel}
M.~R. Akdeniz, Y.~Liu, M.~K. Samimi, S.~Sun, S.~Rangan, T.~S. Rappaport, and E.~Erkip, ``Millimeter {W}ave {C}hannel {M}odeling and {C}ellular {C}apacity {E}valuation,'' \emph{IEEE J. Sel. Areas Commun.}, vol.~32, no.~6, pp. 1164--1179, June 2014.

\bibitem{lakshmirayleigh}
L.~Sau, P.~Mukherjee, and S.~C. Ghosh, ``Priority-{A}ware {G}rouping-{B}ased {M}ultihop {R}outing {S}cheme for {RIS-A}ssisted {W}ireless {N}etworks,'' \emph{IEEE Trans. Network Sci. Eng.}, vol.~12, no.~2, pp. 1172--1185, Feb. 2025.

\bibitem{haenggi2013stochastic}
M.~Haenggi, \emph{Stochastic geometry for wireless networks}.\hskip 1em plus 0.5em minus 0.4em\relax Cambridge Univ. Press, 2013.

\bibitem{cormen199033}
T.~H. Cormen, C.~E. Leiserson, R.~L. Rivest, and C.~Stein, ``Finding the convex hull,'' \emph{Introduction to Algorithms}, pp. 955--956, 1990.

\bibitem{O’Rourke_1998}
J.~O’Rourke, \emph{Search and Intersection}.\hskip 1em plus 0.5em minus 0.4em\relax Cambridge Univ. Press, 1998, p. 220–293.

\bibitem{Ghosh_2007}
S.~K. Ghosh, \emph{Point Visibility}.\hskip 1em plus 0.5em minus 0.4em\relax Cambridge Univ. Press, 2007, p. 13–45.

\bibitem{deBerg2008}
\BIBentryALTinterwordspacing
\emph{Polygon Triangulation}.\hskip 1em plus 0.5em minus 0.4em\relax Springer Berlin Heidelberg, 2008, pp. 45--61.
\BIBentrySTDinterwordspacing

\bibitem{sau2024drams}
L.~Sau, P.~Mukherjee, and S.~C. Ghosh, ``{DRAMS}: {D}ouble-{RIS} assisted multihop routing scheme for device-to-device communication,'' \emph{Comput. Commun.}, vol. 220, pp. 52--63, 2024.

\end{thebibliography}
\vspace{-2mm}
\end{document}